\documentclass[aps,prl,twocolumn,showpacs]{revtex4}
\usepackage{amssymb}
\usepackage{amsmath}
\usepackage{epsfig}

\begin{document}

\title{Driving light pulses with light in two-level media}
\author { R. Khomeriki${}^{1,2}$,  J. Leon${}^1$}
\affiliation {
(${\ }^1$) Laboratoire de Physique Th\'eorique et Astroparticules \\
  CNRS-IN2P3-UMR5207, Universit\'e Montpellier 2, 34095 Montpellier (France)\\ 
(${\ }^2$)  Physics Department, Tbilisi State University, 0128
  Tbilisi (Georgia)}

\begin{abstract}  A two-level medium, described by the Maxwell-Bloch (MB)
system, is engraved by establishing a standing cavity wave with a linearly
polarized electromagnetic field that drives the medium on both ends. A light
pulse, polarized along the other direction, then scatters the medium and couples
to the cavity standing wave by means of the population inversion density
variations. We demonstrate that control of the applied amplitudes of the grating
field allows to stop the light pulse and to make it move backward (eventually to
drive it freely). A simplified limit model of the MB system with
\textit{variable boundary driving} is obtained as a discrete nonlinear
Schr\"odinger equation with \textit{tunable external potential}. It reproduces
qualitatively the dynamics of the driven light pulse. \end{abstract} 
\pacs{42.50.Gy, 42.65.Re}
\maketitle

\paragraph{Introduction.}

Manipulation of light with light has become one of the hottest research spots in
quantum optics this last decade. A widely studied field of research makes use of
electromagnetically induced transparency in three-level systems, which allows to
slow down, and eventually stop, a light pulse \cite{liu,ludkin,bajcsy}. Another
intersting research option uses \textit{resonantly absorbing Bragg reflectors}
(RABR) which consist in a periodic array of dielectric films separated by layers
of a two-level medium \cite{kozekin,malomed,sjohn}, allowing a light pulse not
only to be stopped and trapped \cite{xiao}, but also to be released by
scattering with another control pulse, thus creating a {\em ``gap soliton
memory''} \cite{melnikov}. 

The fundamental process underlying such novel light pulse dynamics is the
cooperative action of nonlinearity and periodicity. Still, a serious drawback
when making use of RABR is the built-in periodic structure that restricts
both the freedom of pulse parameter and of pulse dynamics.

\begin{figure}[ht] \centerline {\epsfig{file=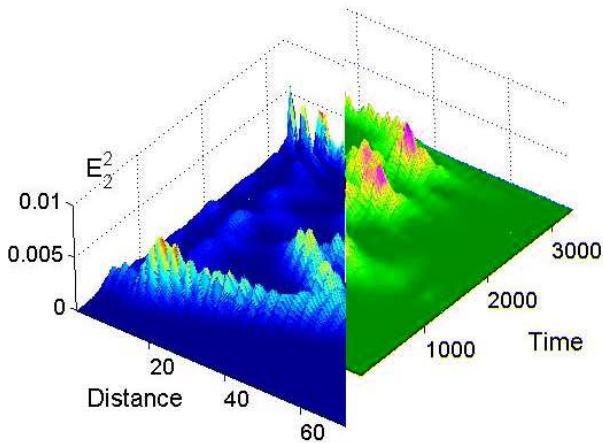,angle=90,width=\linewidth}}
\caption{(Color in line) A typical result of a numerical simulation of
(\ref{MB}) under boundary values (\ref{left-bound}-\ref{right-bound}) with
parameters ${\cal E}_1^{(0)}={\cal E}_1^{(L)}=0.4$, $\omega_1=1.5$, 
${\cal E}_2^{max}=0.04$, $\omega_2=1.31$, $t_0=1000$, $\mu=1/150$. The plot
shows the evolution of the energy density along the polarization direction $y$,
trapped by the grating induced by the stationary wave. The  pulse is eventually
released back at time $4500$ by increasing the amplitude on the incident side
${\cal E}_1^{(0)}$ by 25\%.}
\label{fig:surf1} \end{figure}

We propose to prepare a two-level system (TLS) by establishing a standing
electromagnetic wave in a given polarization direction, and then to scatter a
light pulse, orthogonally polarized. The incident pulse then feels the
\textit{electromagnetic induced grating} through the coupling mediated by the
population density, as described by the governing Maxwell-Bloch (MB) system
\cite{allen,panput}. The freedom in the choice of the standing wave parameters
(in particular the boundary amplitudes) allows us to demonstrate by numerical
simulations as in Fig.\ref{fig:surf1} that the incident light pulse  not only
can be stopped but also can be \textit{released back} to the incoming end.
Engraving a medium with a cavity standing wave is a method previously used to
create two dimensional waveguide arrays in strongly anisotropic photonic
crystals \cite{segev}. 

In a TLS of transition frequency $\Omega$, the MB system is considered in the
isotropic case for a plane polarized electromagnetic field propagating in
direction $z$. The time is scaled to the inverse transition frequency
$\Omega^{-1}$, the space $z$ to the length $\Omega c/\eta$ ($\eta$ is the
optical index of the medium),  the population inversion to the density of active
dipoles $N_0$, the energy to the average $W_0=N_0\hbar\Omega/2$ , the
electric field to $\sqrt{W_0/\epsilon}$ and the polarization to
$\sqrt{\epsilon W_0}$. The resulting dimensionless MB system then reads
\begin{align}
& \textbf{E}_{tt}-\textbf{E}_{zz}+\textbf{P}_{tt}=-\gamma
\textbf{E}_t,\nonumber\\
& \textbf{P}_{tt}+\textbf{P}+\alpha N\textbf{E}=-\gamma_2\textbf{P}_t,
\label{MB} \\
& N_t-\textbf{E}\cdot \textbf{P}_t=-\gamma_1(1+N). \nonumber
\end{align}
where $\textbf{E}$ and $\textbf{P}$, denote vectors in the tranverse plane, e.g.
$\textbf{E}=\left( E_x(z,t),\,E_y(z,t)\right)$. The coupling  eventually results
in a unique dimensionless fundamental constant
$\alpha=2\mu_0c^2/(3\eta^2)|\mu_{12}|^2N_0/(\hbar\Omega)(\eta^2+2)^2/9$ 
(where the dipole moment $\mu_{12}$ is averaged over the orientations
\cite{panput}) and by the normalization of the population inversion density:
$N=-1$ when all active atoms are in the fundamental state, $N=1$  in the excited
state. The dimensionless dissipation coefficients are $\gamma=(c/\eta)({\cal
A}/\Omega)$ resulting from the electric field attenuation $\cal A$, then
$\gamma_1=1/(\Omega T_1)$ and $\gamma_2=2/(\Omega T_2)$ resulting respectively
from the population inversion dephasing time $T_1$  and the polarization
dephasing time $T_2$. 

In the strong coupling case $\alpha\sim1$ (dense media), a multiscale
analysis has shown that the model equation for the two directions of
polarization results to be a system of coupled nonlinear Schr\"odinger equations
\cite{gino}. Then reduced to a unique polarization, it allows for slow light
soliton formation out of evanescent incident light \cite{sls}, namely under
irradiation in the forbidden bang gap. This gap results from the linear
dispersion relation of the MB system, for a carrier $\exp[i(\omega t-kz)]$ on a
medium at rest, namely
\begin{equation}\label{disp}
\omega^2(\omega^2-\omega_0^2)=k^2(\omega^2-1),\quad \omega_0^2=1+\alpha,
\end{equation}
The upper edge of the stop gap at  frequency $\omega_0$
corresponds to $k=0$ and $d\omega/dk=0$.

\paragraph{Numerical simulations.}

We first proceed with numerical simulations of the MB equations (\ref{MB})
submitted to the following boundary-value problem
\begin{align}
&\textbf{E}(0,t)=\left( \begin{array}{c}
{\cal E}_1^{(0)}\cos(\omega_1t) \\{\cal E}_2(t)\cos(\omega_2t)
\end{array}\right),\label{left-bound}\\
&\textbf{E}(L,t)=\left( \begin{array}{c}
{\cal E}_1^{(L)}\cos(\omega_1t) \\0
\end{array}\right),\label{right-bound}
\end{align}
where ${\cal E}_2(t)$ is the slowly-varying low-amplitude pulse envelope
\begin{equation}\label{pulse}
 {\cal E}_2(t)=\dfrac{{\cal E}_2^{max}}{\cosh[\mu(t-t_0)]}.
\end{equation}
The carrier frequencies are chosen close to the gap edge $\omega_0$, inside the
passing band for the grating field, in the gap for the incident pulse, namely
\begin{equation}\label{freq}
 \omega_1>\omega_0,\quad \omega_2<\omega_0,\quad |\omega_j-\omega_0|\sim{\cal O}
(\epsilon^2)
\end{equation}
for $j=1,2$, where $\epsilon$ is our small control parameter. We set from now
on $\alpha=1$.

We have chosen damping coefficients such as to reach a fast stabilization of the
lattice created by the stationary driving: in the $x$-direction we have set
$\gamma=\gamma_2=0.01$ and $\gamma_1=0.001$, while in the $y$-direction they
have been set to zero in order to see clearly the process.
Under such boundary data, with the parameter values indicated in the caption,
we obtain the result displayed in Fig.\ref{fig:surf1}. More precisely, the
applied grating field (polarized along $x$) is settled smoothly and, at time
$t=1000$ the incident pulse enters the medium. It is naturally
trapped and seen to oscillate about the center $z=30$. At time $t=4500$ the
left-hand-side amplitude ${\cal E}_1^{(0)}$ is increased from $0.4$ to $0.5$, 
which produces the reverse motion of the stored pulse. Actualy one may paly with
the driving amplitudes ${\cal E}_1^{(0)}$ and ${\cal E}_1^{(L)}$ to drive the
pulse back and forth.

\paragraph{Interpretation.}

The process described above is now understood, within a multiscale analysis of
MB equations, in terms of a discrete nonlinear Schr\"odiger model with variable
coefficients related to the variation of the boundary grating field amplitudes.
Thanks to (\ref{freq}), a solution of (\ref{MB}) under boundary values
(\ref{left-bound}-\ref{right-bound}) can be sought under the form
\begin{equation}\label{multiscale}
 \textbf{E}(z,t)=\left( \begin{array}{c}
\epsilon E_1(\zeta,\tau_1) \\ 
\epsilon^2 E_2(\zeta,\tau_1,\tau_2)
\end{array}\right)e^{i\omega_0t}+c.c.
\end{equation}
with the slow variables $\zeta=\epsilon z$ and $\tau_n=\epsilon^{2n}t$. The
second slow time $\tau_2$ is meant to capture the nonlinear dynamics of the
low-amplitude ($\epsilon^2$) long duration ($\epsilon^{-2}$) incident pulse.
Note that form assumption (\ref{freq}), the frequency shift from $\omega_0$ is
contained in the slow time variations with $\tau_1=\epsilon^2 t$.

Inserting then (\ref{multiscale}) in the MB system (\ref{MB}) we
eventually obtain (with vanishing damping) 
\begin{align}
&i\frac{\partial E_1}{\partial\tau_1}
-\frac{\omega_0^2-1}{2\omega_0^3}\frac{\partial^2E_1}{\partial\zeta^2}
=\frac{\omega_0^2+3}{4\omega_0}|E_1|^2E_1,\label{evolE1}\\
&i\frac{\partial E_2}{\partial\tau_1}-
\frac{\omega_0^2-1}{2\omega_0^3}\frac{\partial^2E_2}{\partial\zeta^2}
-\frac{\omega_0^2+1}{2\omega_0}|E_1|^2E_2+\nonumber\\ 
&\frac{\omega_0^2-1}{4\omega_0} E_1^2E_2^*
=\epsilon^2\left(-i\frac{\partial E_2}{\partial\tau_2}
+\frac{\omega_0^2+3}{4\omega_0}| E_2|^2 E_2\right),\label{evolE2}
\end{align}
where \textit{star} means complex conjugation. The boundary values
(\ref{left-bound}-\ref{right-bound}) for the $x$-component $E_1$ imply from
equation (\ref{evolE1}) that $E_1(\zeta,\tau_1)$ is a periodic stationary
solution of the nonlinear Schr\"odinger equation with frequency
$\nu_1=(\omega_1-\omega_0)\epsilon^{-2}$. It acts then as an external periodic
potential in the evolution (\ref{evolE2}).

In order to take into account variations $|E_1|(\zeta)$ resulting from the
variations of the boundary driving (\ref{left-bound})
we set (remember $\nu_1\tau_1=(\omega_1-\omega_0)t$)
\begin{equation}
E_1=|E_1|(\zeta)\ e^{i\nu_1\tau_1},\quad
|E_1|^2(\zeta)=V_0(\zeta)+\epsilon^2V(\zeta)
\end{equation}
where $V_0(\zeta)$ is purely periodic while $V(\zeta)$ describes the
\textit{aperiodic inhomegenities} of $|E_1|(\zeta)$ induced by the boundary
values.

We then seek a solution of (\ref{evolE2}) on a suitable orthonormal basis of
Wannier functions $\varphi_j(\zeta)$  which are localized with
respect to the site $j$ \cite{kohn}, within the \textit{one-band approximation}
\cite{wannier-nls}, where $j$ actually indexes the minima of the periodic
potential $V_0(\zeta)$. A solution of (\ref{evolE2}) is sought as
\begin{equation}\label{exp-wannier}
 E_2=\sum_je^{i\nu_1\tau_1}{\cal F}_j (\tau_1,\tau_2,) \varphi_j(\zeta),
\end{equation}
and the equation for the coefficients ${\cal F}_j$ is worked out by inserting
(\ref{exp-wannier}) in (\ref{evolE2}) and by projecting on a chosen $\varphi_j$.
In the \textit{tight-binding approximation} \cite{smerzi} we eventually obtain
\begin{align}
i\frac{\partial{\cal F}_j}{\partial \tau_1}& -(\Omega_0+\nu_1)
{\cal F}_j+\Lambda_0{\cal F}_j^*=
\epsilon^2\big[-i\frac{\partial{\cal F}_j}{\partial \tau_2}+
\Omega_j{\cal F}_j\nonumber\\
&-\Lambda_j{\cal F}_j^*+
Q({\cal F}_{j-1}+{\cal F}_{j+1})+
U|{\cal F}_j|^2{\cal F}_j\big],\label{Fj}
\end{align}
where the coefficients are given from the Wannier basis by
(integrals run on $\zeta\in{\mathbb R}$)
\begin{align*}
&\Omega_0=\frac{\omega_0^2-1}{2\omega_0^3}\int\varphi_j''\varphi_j
+\frac{\omega_0^2+1}{2\omega_0}\int V_0\varphi_j^2,\\
&\Lambda_0=\frac{\omega_0^2-1}{4\omega_0}\int V_0\varphi_j^2,\\
&\Omega_j=\frac{\omega_0^2+1}{2\omega_0}\int V\varphi_j^2,\quad 
\Lambda_j=\frac{\omega_0^2-1}{4\omega_0}\int V\varphi_j^2,\\
&U=\frac{\omega_0^2+3}{4\omega_0}\int \varphi_j^4,\quad 
\epsilon^2Q=
\frac{\omega_0^2-1}{2\omega_0^3}\int\varphi_{j\pm1}''\varphi_j.
\end{align*}
Note that translational invariance guarantees that the above coefficients are
$j$-independent, except of course for $\Omega_j$ and $\Lambda_j$ that bear the
aperiodic inhomogeneity of the external potential $|E_1|(\zeta)$.

Equation (\ref{Fj}) can now be solved first for the $\tau_1$-dependence of
${\cal F}_j$ as a linear system, which provides then a discrete nonlinear
Schr\"odinger coupled system for the $\tau_2$-dependent amplitudes. 
This is done by seeking a solution under the form
\begin{equation}
{\cal F}={\cal G}_j^1e^{-i\Delta\tau_1}+{\cal G}^{-1}_je^{i\Delta\tau_1}
\label{UU}
\end{equation}
for which  the leading order of (\ref{Fj}) furnishes the two coupled linear
equations
\begin{align*}
&-\Bigl[\Omega_0+\nu_1-\Delta\Bigr]{\cal G}_j^1+
\Lambda_0\bigl({\cal G}_j^{-1}\bigr)^*=0,  \\
&-\Bigl[\Omega_0+\nu_1+\Delta \Bigr] \bigl({\cal G}_j^{-1}\bigr)^* +
\Lambda_0{\cal G}_j^1=0,\label{qq}
\end{align*}
The dispersion relation and the relation between ${\cal G}_j^{1}$ and 
$\bigl({\cal G}_j^{-1}\bigr)^*$ automatically follows as
\begin{equation*}
\Delta=\sqrt{\bigl[\Omega_0+\nu_1\bigr]^2-\Lambda_0^2}, \qquad 
\bigl({\cal G}_j^{-1}\bigr)^*=\frac{\Lambda_0}{\Omega_0+\nu_1+
\Delta}{\cal G}_j^{1}.
\end{equation*}
At next order we readily get for ${\cal G}_j^1$ the discrete nonlinear
Schr\"odinger equation, the parameters of which are greatly simplified if
we note that in our numerical simulations (for sufficiently deep lattice)
$\Omega_0\gg \Lambda_0$ and $\Omega_j\gg\Lambda_j$. In such a case 
the equation reads
\begin{equation}
i\frac{\partial{\cal G}_j^1}{\partial\tau_2}-\Omega_j{\cal G}_j^1
=Q[{\cal G}_{j+1}^1+{\cal G}_{j-1}^1]
+U|{\cal G}_j^1|^2 {\cal G}_j^1.\label{DNLS}
\end{equation}
The electric field envelope in the $y$-direction of polarization reads
\begin{equation}\label{solution}
 E_2=\sum_je^{-i\Omega_0\tau_1}{\cal G}_j^1 (\tau_2) \varphi_j(\zeta),
\end{equation}
in terms of the solution of the chosen Wannier basis.

The dynamics of the pulse is thus interpreted out of the discrete nonlinear
Schr\"odinger model (\ref{DNLS}) as the action of the \textit{potential}
$\Omega_j$ that translates the applied variations of the boundary driving in the
$x$-direction of polarization.

\paragraph{Application.}

\begin{figure}[ht]
\centerline{\epsfig{file=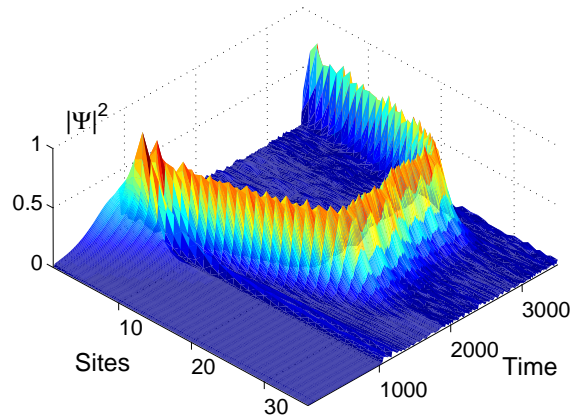,width=\linewidth}}
\caption {(Color in line) Plot of the solution of (\ref{DNLS}) with the
parameters $Q=0.2$, $U=0.4$ and an incident pulse ${\cal
G}_0^1(t)=0.305e^{-0.418\,i t}/\cosh[(t-750)/250]$ which propagates, becomes
trapped and is then released back by the action of $\Omega_j$ in
(\ref{pot}).\label{fig:surf2}}\end{figure}

In order to illustrate the above interpretation, we proceed now with numerical
simulations of (\ref{DNLS}) where $\tau_2=t$ to read quantities in physical
dimensions. The potential $\Omega_j$ models the variations of $|E_1|$ away from
a purely periodic function as soon as ${\cal E}_1^{(0)}\ne{\cal E}_1^{(L)}$. We
set
\begin{equation}\label{pot}
\Omega_j=\left\{\begin{array}{ccl}0 &,\quad &t<650,\\
 -0.07\ j &,\quad & t>650\end{array}\right.
\end{equation}
and obtain the Fig.\ref{fig:surf2} which shows the same qualitative behavior as
Fig.\ref{fig:surf1}

\paragraph{Conclusion and comments.}

We have shown that the Maxwell-Bloch system, the celebrated fundamental and
general semi-classical model of interaction of radiation with matter, may serve
as a tool to store and drive light pulses in an arbitrary way by conveniently
using laser light to engrave the medium with a controllable standing wave
pattern. The extreme genericity of MB model together with the freedom in the
boundary values of the engraving field, constitute a decisive advantage over
other techniques to store and manipulate light pulses.

The process is understood by deriving a discrete nonlinear Schr\"odinger model
where the external tunable potential actually translates the effects of boundary
driving variations. Our purpose was simply to provide a \textit{qualitative}
interpretation. 

A more detailed study, reported to future work, would require first
to use exact solutions of (\ref{evolE1}) for the grating field (e.g. in terms of
Jacobi elliptic functions), second to construct the corresponding most adequate
Wannier basis, third to evalute precisely the effect of boundary driving
variations on the factor $\Omega_j$, and last to study the full system
(\ref{Fj}) that couples ${\cal F}_j$ to ${\cal F}_j^*$.

\paragraph{Acknowledgements.} Work done under contract CNRS GDR-PhoNoMi2
(\textit{Photonique Nonlin\'eaire et Milieux Microstructur\'es}). R.K.
aknowledges invitation and support of the \textit{Laboratoire de Physique
Th\'eorique et Astroparticules} and USA CRDF award \# GEP2-2848-TB-06.

\end{document}